\documentclass{article}
\usepackage{authblk}
\usepackage[a4paper, margin=1in]{geometry}
\usepackage[utf8]{inputenc}
\usepackage[T1]{fontenc}
\usepackage[numbers]{natbib}
\usepackage{pgfplots}
\usepackage[a4paper, margin=1in]{geometry}
\usepackage[hyphens]{url}
\usepackage[english]{babel}
\usepackage{caption}
\usepackage{pgf-pie}
\usepackage{tikz}
\usepackage{pgfplots, pgfplotstable}

\usetikzlibrary{arrows, positioning, shadows}
\definecolor{myred}{HTML}{ef476f}
\definecolor{myblue}{HTML}{118ab2}
\definecolor{myyellow}{HTML}{ffd166}
\definecolor{mygreen}{HTML}{06d6a0}
\definecolor{mydarkblue}{HTML}{073b4c}
\usepackage{amsmath}
\usepackage{graphicx}
\usepackage{todonotes}
\usepackage{caption}
\usepackage{subcaption}

\title{Hello Me, Meet the Real Me: Audio Deepfake Attacks on Voice Assistants}

\begin{document}
\author[1]{Domna Bilika}
\author[1]{Nikoletta Michopoulou}
\author[1]{Efthimios Alepis}
\author[1,2]{Constantinos Patsakis}

\affil[1]{Department of Informatics, University of Piraeus, 80 Karaoli \& Dimitriou str., 18534 Piraeus, Greece}

\affil[2]{Information Management Systems Institute of Athena Research Centre, Greece}

\date{}
\maketitle

\begin{abstract}
The radical advances in telecommunications and computer science have enabled a myriad of applications and novel seamless interaction with computing interfaces. Voice Assistants (VAs) have become a norm for smartphones, and millions of VAs incorporated in smart devices are used to control these devices in the smart home context. Previous research has shown that they are prone to attacks, leading vendors to countermeasures. One of these measures is to allow only a specific individual, the device's owner, to perform possibly dangerous tasks, that is, tasks that may disclose personal information, involve monetary transactions etc. To understand the extent to which VAs provide the necessary protection to their users, we experimented with two of the most widely used VAs, which the participants trained. We then utilised voice synthesis using samples provided by participants to synthesise commands that were used to trigger the corresponding VA and perform a dangerous task. Our extensive results showed that more than 30\% of our deepfake attacks were successful and that there was at least one successful attack for more than half of the participants. Moreover, they illustrate statistically significant variation among vendors and, in one case, even gender bias. The outcomes are rather alarming and require the deployment of further countermeasures to prevent exploitation, as the number of VAs in use is currently comparable to the world population.
\end{abstract}


\section{Introduction} 
Digital Assistants (DAs), also referred to as Virtual Assistants, Intelligent Personal Assistants and Artificial Intelligence Assistants, are used more and more as their sophistication and capabilities are rapidly increasing, with the manufacturing of new products and services on top of them. DAs are defined as devices - usually speakers - or services integrated into mobile phones and web services that use advanced artificial intelligence (AI) and other advanced algorithmic approaches to i) perform tasks for an individual, ii) answer various questions, iii) maintain a conversation with the user and iv) retain information about the user and issue reminders and warnings based on environmental constraints, e.g., time and location. The above makes DAs extremely useful for people with mobility problems and the elderly. Chambers and Beaney \cite{ruth2020potential} describe how VAs can be applied to patients’ health and care needs. Also, Pradhan et al. \cite{alisha2020use} conducted an experiment with VAs over a period of 3 weeks on older people who do not use computing devices every day. In the end, elderly people consistently use the device to access information online, mainly to access health-related information.

DAs use natural language processing (NLP), natural language understanding, and machine learning to learn and provide a personalised conversational experience continuously. Combining historical information such as purchase preferences, home ownership, location, family size, and so on, the underlying algorithms can create data models that identify behavioural patterns and then refine those patterns as data is added. By learning users' history, preferences, and other information, DAs can answer complex questions, provide recommendations, make predictions, and even initiate conversations. They facilitate users' daily lives by performing functions to manage electrical appliances, even those related to home security \cite{benjamin2020smart}. DAs are always there for users to inform them of any outstanding work to be done through reminders. A significant benefit is that they can simultaneously serve a large percentage of people. Finally, with prolonged use, they can gather more useful information to improve the user experience.

Depending on the input format, DAs can be classified into three main categories. If their input is textual, we usually refer to them as \textit{chatbots}. When the DA interacts with the user using voice, the DA is referred to as \textit{voice assistant}. Finally, some DAs use visual input like digital images, videos, or a live camera. These assistants have the ability to do image processing to recognise objects in the image to help the users get better results from the clicked images. Computer vision also enables the system to recognise body language, which is a significant part of communication.

To provide authentication followed by access to sensitive data, VAs can be trained to "obey" only one user or allow this single user to perform sensitive tasks, e.g., use services that have payments. Every VA requires user training before using it for the first time. Due to the increased volume of data they receive daily, they are becoming more and more efficient. However, training in a single voice is still quite challenging. It is worth noting that not all voice commands work in the same way since there are commands that are executed by everybody, including the voice of the user who `trained' the VA. Still, some commands must have been explicitly uttered by the user who trained this assistant, as they are considered dangerous, e.g., monetary transactions, phone calls and reading messages.

As VAs become more popular, there are increasing security, privacy, and legal risks involved. This study aims to attack VAs trying to bypass the aforementioned restrictions. For this purpose, we collect voice data from different sources of the \textit{trusted} by the VA voice to create voice samples that would trigger voice commands that are restricted. Thus, we examined ways in which we can extract information about a user through a number of available resources. For instance, we used face-to-face recordings and videos as input for our experiments. Another way of collecting data would be via phone calls. Nevertheless, it was inapplicable for most of our participants, so it has been excluded from our experiments. As a second step, following the data collection, voice synthesising with a third-party application took place in our experiments. Finally, the produced synthesised output was played from another device to attack the VA with appropriate commands. To do this, there are many methods, e.g., the adversary plays the audio when in proximity to the VA, or the audio is reproduced by the smartphone, triggering the VA \cite{alepis2017monkey,zhang2018using} exploiting the fact that some VAs do not distinguish the source of the audio. 

\noindent\textbf{Scope:} With the continuous integration of VAs in several devices in our homes, several devices are waiting to collect voice commands. As already discussed, some of these commands may have a significant impact on the users, e.g., sensitive information leakage and financial cost. To prevent such attacks, manufacturers have installed specific features that may allow only `trusted' voices to perform such sensitive tasks. This begs the question of assessing the trustworthiness of such a protection mechanism. To answer this, we have to consider it in the context of voice synthesis and the wide availability of voice samples of users in modern societies. The latter is a crucial parameter, as in the ubiquitous computing environment that we are living in, a plethora of means can record one's voice. 

Therefore, the main goal of this research is to determine whether an adversary can replicate a VA's `trusted' voice in a real-world setting. To this end, we conduct a set of targeted experiments that harness the voice of a user from various sources to train a voice synthesis model and use it to issue a sensitive command against two of the most widely used proprietary VAs, namely Google's Assistant and Apple's Siri. As a next step, we use the same trained model to attack voice authentication systems to determine whether commercial off-the-shelf systems that use such authentication are vulnerable to such attacks. To the best of our knowledge, this is the first work in the literature to perform a deepfake audio experiment in a broad and open setting, shedding light on the security of a technology that is continuously being integrated into devices and services.

\noindent\textbf{Main contributions and results:} Our results illustrate a rather alarming state of practice in two of the most used VAs we tested. In practice, we show that around 3 out of 10 of our attacks successfully deceive the VAs into performing an action that should be performed only by authorised users, using an off-the-shelf open-source solution. Moreover, our research indicates that these results may significantly vary among vendors and even gender. Indeed, in our experiments, measurements between vendors illustrate huge differences in users' exposure. At the same time, for one OS, it is shown that there is a similar gap which depends on the gender of the simulated voice. The latter practically illustrates the possible gender biases in cybersecurity research.

\begin{figure*}
    \centering
    \includegraphics[width=\textwidth]{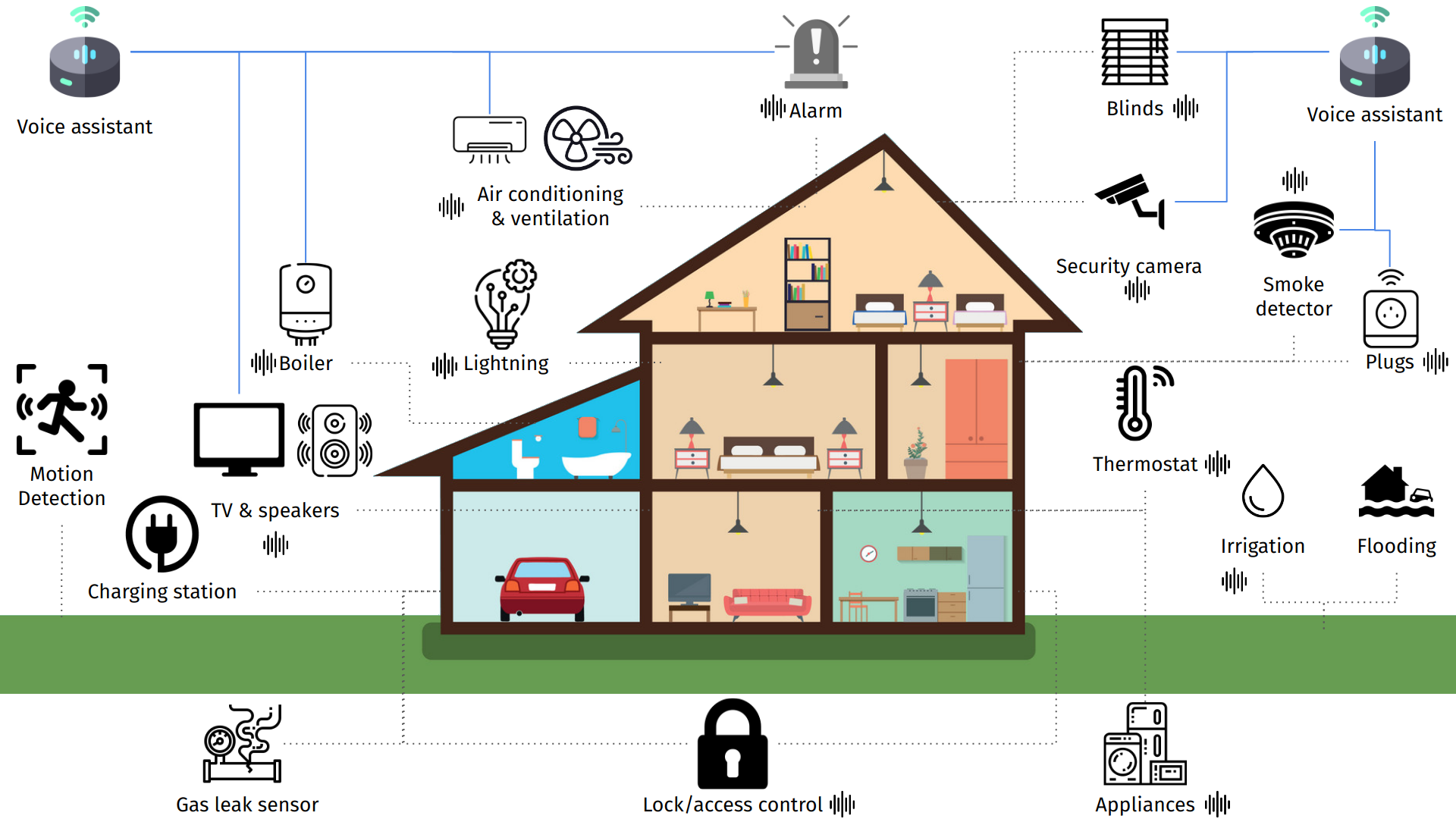}
    \caption{Devices and sensors in a Smart Home. The voice icon (\includegraphics[width=0.1in]{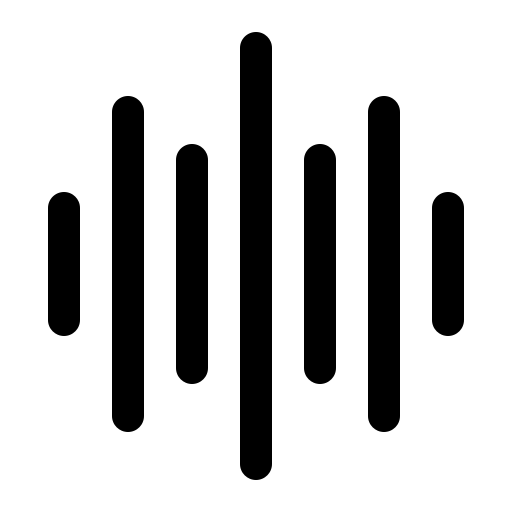}) indicates devices controlled by VAs.}
    \label{fig:smart_home_voice}
\end{figure*}

\noindent\textbf{Potential impact:}
In 2019, approximately 3.25 billion VA devices were purchased around the world. Forecasts suggest that by the end of 2023, the number of VAs will reach around 8 billion units – a number in the scale of the world’s population \cite{chen2023survey}. VAs are a feature found in many consumer electronics devices, ranging from smartphones to mobile-operated car systems. Thus, it is essential to  understand the extent of the risks involved in attacks on VAs. Notably, VAs are usually not located in isolated environments or only as smartphone apps. A quite typical case where VAs are increasingly being used is a Smart Home. This term refers to an integrated system of interconnected devices, sensors, and services which automates various tasks inside a house and can be automatically controlled remotely from Internet-connected devices such as smartphones and tablets. The user can remotely control and schedule functions such as access to the house and premises, activation and deactivation of a device, control of the alarm system etc. These devices are usually connected to a central “gateway”. This way, the user can control all the connected appliances, including but not limited to lighting, thermostat, boilers, and so many other functions through a personal device, even if they are physically far away from it. At any time, they are aware of any operation of the house through relevant notifications. However, there are many risks, mainly regarding security issues affecting users. Given that VAs are often used to control home automation, if they can easily be deceived, the impact on home automation can be catastrophic. Beyond electricity costs, a Smart Home also allows for physical access automation, implying that attacks may extend the cyber layer and reach the physical one. 
In Figure \ref{fig:smart_home_voice}, we try to illustrate the devices and sensors indicating the ones which a VA can control to understand the potential risks from their abuse.

\noindent\textbf{Ethics:} To perform this study, an extensive experiment was conducted with 140 people participating on a volunteer basis. We detailed the scope, goals, and steps of the experiment to all participants before their participation. All steps were performed by the participants without installing anything on their devices. Additionally, in the attacks, the participants did not have any costs.

\noindent\textbf{Road map:} The rest of this work is organised as follows. In the next section, we analyse the related work regarding VAs and voice synthesis. In Section \ref{sec:sources}, we describe how data can be collected from various media and manipulated accordingly. We describe data collection with malicious software and proceed with the processing and synthesis of voice data. The two most important sections are attacks on VAs, which are demonstrated through our application showing how easily a system can be tricked with a synthesised voice into performing protected actions. The robustness and validity of our results are guaranteed through the experimental results of a large group of people that participated in our research. Finally, this concludes by summarising our contributions and discussing possible countermeasures and future work.

\section{Related Work}
\label{sec:related}
In the following paragraphs, we first provide an overview of the attacks on VAs, trying to illustrate the various risks users are exposed to. Then we present how voice synthesis works, focusing on the methods employed by the tool we used to synthesise commands with the participants' voices.

\subsection{Attacks on Voice Assistants}
Kumar et al. \cite{kumar2018skill,kumar2019emerging} introduced the skill squatting attack targeted at Alexa. In essence, the researchers used prerecorded samples to identify where the VA misinterprets the audio input. Some of these errors were found to be consistent, so they could be exploited to lure a user to a malicious application without being aware of it. The attack could be further tuned into a spear phishing attack to target specific demographic groups.

In the REEVE attack \cite{yuan2018all} Yuan et al., the attacker performs radio signal injection to trick Alexa in Amazon's Echo into performing specific tasks, identifying more than one hundred vulnerable skills and applets.

A critical issue in VAs is the fact that humans and machines conceive audio totally differently. The latter opens the door to a wide range of attacks \cite{191968}, and wrong triggers \cite{dubois2020speakers} as what we hear and say is not translated the same way in a machine. Moreover, there is a wide audio spectrum that could be sensed by a machine but not by the human ear. As a result, an adversary can generate audio that does not make any sense for a human but is interpretable by a machine \cite{carlini2016hidden}. Practically, the adversary can use any device to play the audio which could trigger a command to a VA. Such audio may even be embedded in songs \cite{217607}, making it impossible for the human to understand the origin of the attack. Similar attacks can be considered the DolphinAttack \cite{zhang2017dolphinattack} and the ones from Sch{\"o}nherr et al. \cite{schonherr2018adversarial}, Roy et al. \cite{211283} and Yan et al. \cite{yan2020surfingattack} as the audio is inaudible to the human ear but interpretable by a machine.

In general, the apps and skills of VAs are not properly monitored, allowing malicious ones to appear in the relevant stores \cite{cheng2020dangerous,lentzsch2021hey}. In this context, Zhang et al. \cite{zhang2019dangerous} introduced two new attacks, namely voice squatting and voice masquerading. In the former, the attacker uses similarly pronounced phrases to trigger malicious skills instead of benign ones. In the voice masquerading attack, the adversary impersonates the voice of the VA or of another legitimate skill to lure the user into disclosing sensitive information.

Attempting to infringe on the Android operating system is not an emerging research object. In every version of Android, attempts have been made to find vulnerabilities. Diao et al. \cite{diao2014your} presented an approach (GVS-Attack) to launch permission bypassing attacks from a zero-permission Android application (VoicEmployer) through the speaker. Through the Android Intent mechanism, VoicEmployer triggers Google Voice Search in the foreground and then plays prepared audio files in the background. Google Voice Search can recognize this voice command and execute corresponding operations. Also, they found a vulnerability in status checking in the Google Search application, which can be utilised to dial arbitrary numbers even when the phone is securely locked with a password.

Alepis and Patsakis \cite{alepis2017monkey} thoroughly examine in their research the dangers lurking in mobiles with intelligent VAs. They deny that it is a fictitious threat but a real scenario that can greatly expose users. In this work, detailed real scenarios of attacks with voice commands were implemented. However, due to the use of AI, these systems are more difficult to break. An important point of their research is the fact that attacks on mobile devices were not limited. There is a variety of different devices that incorporate VAs such as smartwatches, personal computers, or even smart TVs.

In another research Zhang et al. \cite{zhang2018using} proposed a stealthy attacking method targeting VAs on smartphones. They proposed an attacking method that could activate the VA and apply further attacks, such as leaking private information, sending forged SMS/Emails, and calling arbitrary numbers. To hide the attack from users, an optimal attacking time was chosen. Through their proof-of-concept attack targeting Google Assistant on the Android platform, they demonstrated the feasibility of the attack in real-world scenarios.

Esposito et al. \cite{10.1145/3488932.3497766} recently took advantage of Alexa's inability to distinguish voice commands from audio files that it reproduces to trick the VA into performing unauthorised actions. Similarly, Vaspy of Zhang et al. \cite{zhang2019activated} is a spyware app which records activation keywords and replays in targetted intervals  to launch an attack on the VA locally, solely studying the case of a handful of Android phones. Chen et al. \cite{chen2020devil} synthesise voice commands that are not interpretable by most humans yet sound like normal speech to launch attacks against commercial VAs. However, their attacks target generic automatic speech recognition systems and do not cover the cases of trusted voices. Finally, Wenger et al. \cite{wenger2021hello} used existing voice datasets to train commercial speaker recognition and then used voice synthesisers to attack them with high success rates.

For more on the security of VAs, the interested reader may refer to \cite{abdullah2020sok,edu2020smart,bolton2021security,yan2022survey}.

\subsection{Voice synthesis}
For centuries people have been experimenting with the development of devices that could replicate and produce the voice of animals and, ultimately, humans. Nevertheless, the first devices to produce human speech date back to the 18th century with the pioneering works of Kratzenstein and von Kempelen. With the technological advances, researchers managed not only to replicate the human voice but to synthesise voices that can even utter arbitrary texts. To achieve this, there are several approaches. For instance, we have articulatory synthesis \cite{coker1976model} where, as the name implies, the goal is to simulate the mechanism that humans speak; thus, one simulates the lips, tongue etc. There is also concatenative synthesis \cite{moulines1990pitch} in which the goal is first to isolate various speech segments, ranging from sentences down to diphones and phonemes, and use this as a reference to transform text to voice. In formant synthesis, one tries to replicate the  pitch (frequency) and volume (amplitude) of a voice, following the way that we replicate the music produced by musical instruments \cite{klatt1980software}. Statistical parametric speech synthesis \cite{yoshimura1999simultaneous} uses a statistical model to generate speech. The model is trained on a large dataset of speech samples to learn the statistical patterns that govern how sounds are produced. Once the model is trained, it can be used to generate new speech samples by sampling from the learned distribution of sounds. This can be done by specifying the desired text to be spoken, along with any desired prosodic features such as pitch and speaking rate. The model then generates a synthetic speech sample corresponding to the specified text and prosodic features. There are several different types of statistical parametric speech synthesis models, including hidden Markov models and neural network-based models. Finally, we have neural speech synthesis, which started as a type of statistical parametric speech synthesis using neural networks as the underlying model. The basic idea is to use a neural network to map text input to speech output. The network is trained on a large dataset of speech samples to generate the corresponding speech waveform.

Different types of neural networks can be used for speech synthesis, including feedforward networks and recurrent networks. Recurrent networks are particularly popular because they can handle sequential data, such as speech, which has temporal dependencies. In a neural speech synthesis system, the input text is first processed by a text encoder network that maps the text to a high-dimensional representation. This representation is then passed to a speech decoder network that generates the speech waveform. The decoder network is typically a generative model, such as a Generative Adversarial Network (GAN) or a Variational Autoencoder (VAE), which generates the final speech output. This method of speech synthesis has been shown to produce high-quality speech that is very similar to human speech. Additionally, neural speech synthesis can also be used to control various aspects of the speech output, such as speaking rate and pitch, by conditioning the network on these features during training.

For the composition of a synthesised user's voice that we utilised in the attacks on VAs, we used the ``Real Time Voice Cloning'' (RTVC) \cite{jemine2019master}, which relies on the work of Jia et al. \cite{jia2018transfer}. It is a three-stage deep learning framework that performs voice cloning in real time. Using an utterance of speech of 5 seconds, the framework can capture in a digital format a meaningful representation of the spoken voice. Thus, by giving a text prompt, it can perform the text-to-speech conversion using any voice extracted by this process. After long hours of training and by using a large dataset, the framework could clone voices it has never heard of and generate speech from arbitrary text. Since it performs neural speech synthesis, the application consists of three parts. The first part is a speaker encoder that derives an embedding from the short utterance of a single speaker. The embedding is a meaningful representation of the voice of the speaker. The second part is a synthesizer that, conditioned on the embedding of a speaker, generates a spectrogram from text and the last one is a vocoder that infers an audio waveform from the spectrograms generated by the synthesizer.

More precisely, RTVC uses the synthesizer Tacotron \cite{WangSSWWJYXCBLA17}, which is a recurrent sequence-to-sequence model that predicts a mel spectrogram from the text. RTVC features an encoder-decoder structure that is bridged by a location-sensitive attention mechanism. Individual characters from the text sequence are first embedded as vectors. Convolutional layers follow to increase the span of a single encoder frame. These frames are passed through a bidirectional Long Short-Term Memory (LSTM) to produce the encoder output frames. This is where SV2TTS (Transfer Learning from Speaker Verification to Multispeaker Text-To-Speech Synthesis) \cite{jia2018transfer} modifies the architecture: a speaker embedding is concatenated to every frame that the Tacotron encoder produces. The attention mechanism attends to the encoder output frames to generate the decoder input frames. Each decoder input frame is concatenated with the previous decoder frame output passed through a pre-net, making the model autoregressive. This concatenated vector goes through two unidirectional LSTM layers before being projected to a single mel spectrogram frame. Another projection of the same vector to a scalar allows the network to predict on its own that it should stop generating frames by emitting a value above a certain threshold. The entire sequence of frames is passed through a residual post-net before it becomes the mel spectrogram.

In addition, it is worth mentioning that in SV2TTS and Tacotron, WaveNet is the vocoder \cite{vanwavenet}. WaveNet has been at the heart of deep learning with audio since its release and remains state of the art regarding voice naturalness in Text-to-Speech (TTS). However, it is also known for being the slowest practical deep learning architecture at inference time. More recent research has improved it to make the generation near real-time or faster than real-time without significantly impacting the quality of the generated speech. Thus, in RTVC, WaveRNN \cite{kalchbrenner2018efficient} has been selected.

\section{Attack scenarios}
\subsection{Threat assumptions}
The goal of this work is to assess the security of commercial VAs against voice synthesis attacks. Since these commercial VAs 
are closed-source and little information about their internal mechanisms is provided, we consider them as black boxes. Moreover, we assume an adversary which can collect some voice samples of the victim (see paragraphs below) and has the capacity to get in proximity to the victim's VA unattended and play an audio. For the VA, we assume that it has been trained with the victim's voice to perform potentially dangerous tasks. In our experiments, we consider two of the most widely used VAs, namely Google Assistant and Siri. However, the method is generic enough to attack any other VA. 
\subsection{Possible information sources}
\label{sec:sources}
As mentioned above, the aim of this research is to evaluate attacks on VAs using synthesised voice commands. Clearly, to achieve this, the attacker must collect voice samples which may originate from different sources from the potential victim. We highlight this heterogeneity as the quality of the samples from different sources may vary for various reasons. These may include, among others, the presence of background noise, small-length samples, and poor recording quality. Nevertheless, since we are in the information era where people share everything mainly through the internet, one has to understand that with a large amount of available data on the internet through, e.g., social media, it becomes easier to extract it for attacks. More specifically, we consider four distinct and realistic attack scenarios, which are analysed in the following paragraphs. In three of them, we consider an active adversary who interacts with the potential victim via three different modalities. Finally, in the last scenario, we consider a passive adversary.

\subsubsection{Face to Face user interaction}
The first way of collecting data requires personal contact with the \textit{victim}, the targeted individual whose VA would be attacked. All that is needed is a recording device that can be activated during the conversation with the potential victim. Evidently, an important problem encountered in such cases is the maintenance of favourable conditions in terms of sound, as a lot of noise can distort the collected data.

\subsubsection{Via a Call}
An alternative to collecting the voice samples is via call recording. In this case, it was a less "risky" way as it did not bring the attacker into personal contact with the “victim” and, therefore, could not be perceived by the latter as being recorded. All that was needed was to make a call to the victim under any pretext. This can be considered a social engineering attack as the attacker needs to maintain a conversation with the victim for enough time to collect the necessary voice samples. Evidently, the network quality and the location of the victim play a crucial role, as these two factors can significantly affect the quality of the recording. 

\subsubsection{Spyware and app vulnerabilities}
Numerous malicious applications appear in Google Play and App Store that occasionally appear, many of which manage to trick millions of users into installing them. Moreover, the fact that some of them do not appear to act malicious does not necessarily mean that they are not abusing users' data or that they do not contain vulnerabilities that can be abused, even in the backend. 

Given that numerous applications in Google Play and App Store use the microphone permission to record users' voices, many of which may remotely store the collected information, it is clear that thousands of developers may access the users' voice samples. Should the application be considered trusted by the user, the samples can be ample, guaranteeing that there would be enough without background noise and of good quality to be used for synthesis. Similarly, unprotected AWS buckets and vulnerable APIs may grant an attacker access to backends that apps store video or audio samples.

\subsubsection{Social media}
Technology is increasingly invading people's daily lives. Social media are gaining an indispensable role in our lives. Users publish various images, videos and audio content freely, providing access to them to almost everyone. This content is, in many cases, of very high quality as it is recorded via modern cameras and microphones, which have very high resolution, sensitivity and inherent noise-cancellation mechanisms. Moreover, on many social media, users share their thoughts or speak in interviews, so the background noise is minimal. Therefore, an attacker can easily get hold of ample content and extract many voice samples of a potential victim without the need for direct interaction or consent. 

\subsection{Attack preparation and impact}
Based on the above four scenarios, we can safely assume that after the first steps, all four scenarios lead to the same output, an audio file containing the user's voice. In the first three scenarios, this file is the immediate outcome of the recording. However, in the case of social media, since users most often share videos than audio content, the attacker has to extract the audio from the video. Solutions such as FFmpeg\footnote{\url{https://ffmpeg.org/}} can easily perform this task without degrading or altering the audio quality. Once the adversary has the audio files, their next task is to isolate the victim's voice. For this task, she may use Audacity\footnote{\url{https://www.audacityteam.org/}}, which can automatically split audio files into segments based on silenced parts. Then, these files are sent to the voice synthesiser, which utilises them to create an audio file that, machine-wise, resembles the user's voice. This file is then reproduced to the VA, which would execute a command allowing the adversary to perform a task with elevated privileges, a task that would only be allowed by the trusted voice of the targeted user. Depending on the command, the impact could range from information leakage, e.g., reading emails, messages etc., and money loss, e.g., ordering a product/service using the user's stored credit card, to physical attacks, e.g. granting access to the victim's premises by unlocking a door. Other attacks, such as resource exhaustion and appliance fatigue leading to breakdowns or monetary loss and reputation damage through posting unauthorised content on social media and communication platforms, are also other possible impacts. The attacker could even exploit VAs to collect 2FA tokens, further deepening the possible impact of such an attack.  

\begin{figure*}[!th]
    \centering
    \includegraphics[width=.8\textwidth]{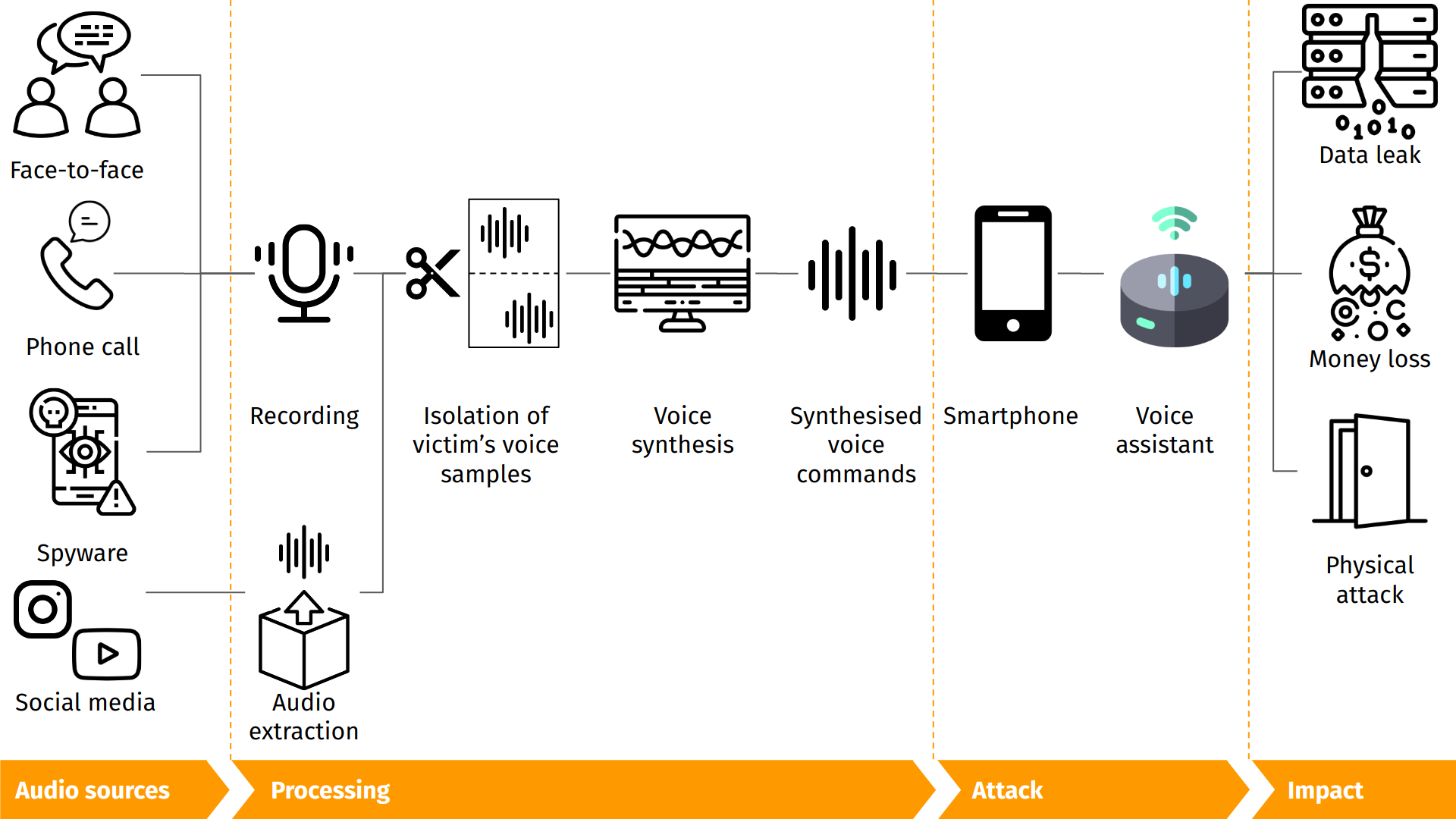}
    \caption{The attack scenarios considered in this work}
    \label{fig:attack_scenarios}
\end{figure*}

In Figure \ref{fig:attack_scenarios}, we illustrate the possible attack scenarios that were described above. Moreover, we illustrate the possible impact.

\subsection{Training the voice assistants and authentication}
Our experimental process took into consideration the training of the VAs from Google and Apple to authenticate the user's voice so that they respond only to the user to which they have been trained. 
In both cases of the examined VAs, Google Assistant and Siri, the mobile operating systems ask the user to speak specific commands, aiming both in the calibration of their speech recognition module and also to recognise the owner of the mobile device successfully. Following this, in our experiments, we have chosen a text containing commonly spoken phrases from users' daily lives. 

The fact that the companies acknowledge that they perform voice authentication is evident, for instance, by Google stating that 
\begin{quote}
    \textit{``When you turn on Voice Match, you can teach Google Assistant to recognize your voice so it can verify who you are before it gives you personal results.''}\footnote{\url{https://support.google.com/assistant/answer/9071681}}
\end{quote}
Similarly, Apple states for iOS that
\begin{quote}
    \textit{``You can control iPhone with just your voice. Speak commands to perform gestures, interact with screen elements, dictate and edit text, and more.''}\footnote{\url{https://support.apple.com/en-gb/guide/iphone/iph2c21a3c88/ios}}
\end{quote}
Moreover, Homepod also uses Siri and the fact that it is trained with a specific voice that it is considered trusted is acknowledged by the following quote:
\begin{quote}
    \textit{``Siri can recognise multiple voices, so everyone in your home can use HomePod to enjoy personalised music recommendations, access their own playlists, send and read messages, make phone calls and more.''
}\footnote{\url{https://support.apple.com/en-gb/guide/homepod/apd1841a8f81/homepod}}
\end{quote}

\section{Experimental results}
To assess the security offered by VAs, we conducted an experiment targeted on Android and iOS, the two most popular mobile Operating Systems, that tries to replicate the attacks that could be launched from face-to-face, spyware, and social media. We did not consider phone call attacks as they could be considered a case of a face-to-face attack with possibly additional noise due to the network. Moreover, they required more interaction with the experiment participants, many of which might not feel comfortable sharing their telephone numbers. In what follows, we detail the experiment and its findings.

\begin{figure*}[!th]
    \centering
    \includegraphics[width=\textwidth]{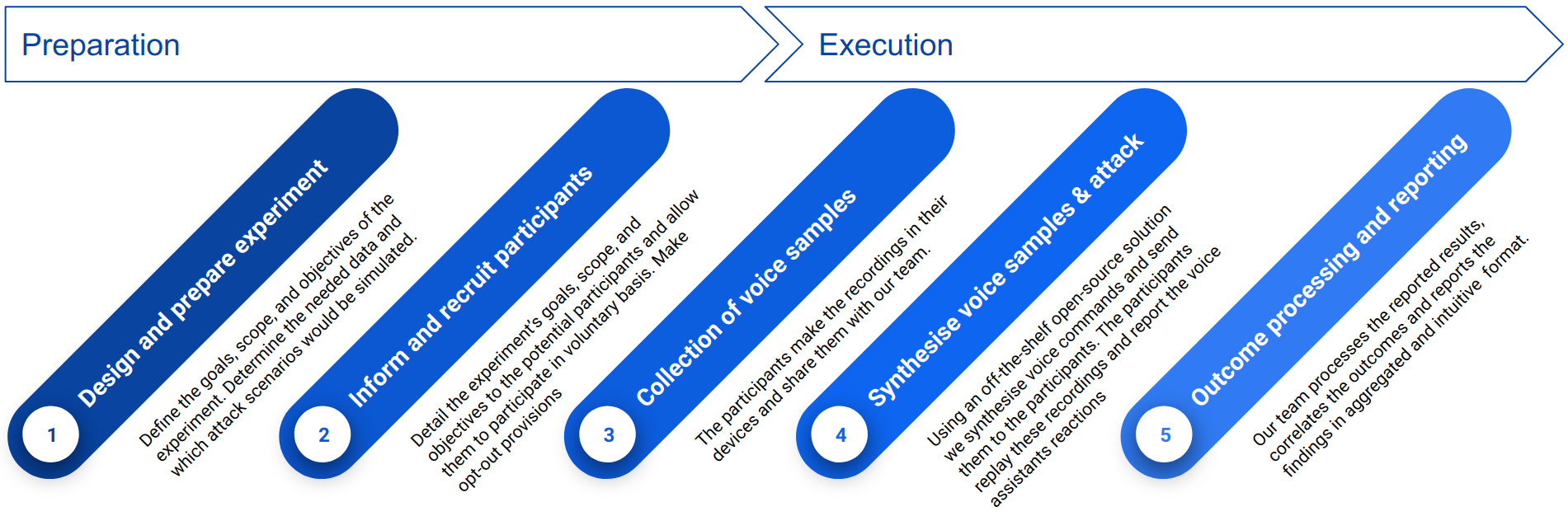}
    \caption{The phases of the experiment.}
    \label{fig:experiment_phases}
\end{figure*}

The phases of the experiment are illustrated in Figure \ref{fig:experiment_phases}.

\subsection{Concept and experimental setup}
To conduct our experiment, we first modelled the attacks and tried to alleviate some technical issues, e.g. splitting the recordings to keep only the victim's voice etc. We concluded that we would request the participants to provide us with clear recordings of the same given text, see Figure \ref{fig:text}. We consider that only two samples are needed, one recorded through a mobile phone and one from another device, e.g. a PC. This way, we would have the voice of the participant from an arbitrary text of a realistic length without requesting the participants to perform exhaustive actions but enough to synthesise their voice. To scale the experiment and in order not to intervene and tamper with the participants' devices, we opted to provide the participants with the synthesised samples and request them to report the actions of their VA. This way, the participants had the necessary guarantees that no harm would be caused to their software and/or hardware and also had full control of opting out. 

Based on the above, we created a description of the experiment and notified potential participants orally and in written form of the scope, goals, and steps of our experiments. Therefore, the participants were fully informed and voluntarily opted in, and, as already mentioned, could opt out whenever they deemed appropriate. Participants who undertook the experiments, after training the VA, emailed the two requested recordings. For each participant, we synthesised some voice samples that would trigger the VA and request a phone call. These recordings were sent to the participants, who replayed the recordings to their VAs, monitored their actions and subsequently sent us back their results for further processing and examination.     

\begin{figure}[!th]
    \centering
    
    \noindent\fbox{%
    \parbox{\columnwidth}{%
        \small 
        \texttt{While eating at a restaurant is an enjoyable and convenient occasional treat, most individuals and families prepare their meals at home. To make breakfast, lunch, and dinner daily, these persons must have the required foods and ingredients on hand and ready to go; foods and ingredients are typically purchased from a grocery store, or an establishment that distributes foods, drinks, household products, and other items that're used by the typical consumer. Produce, or the term used to describe fresh fruits and vegetables, is commonly purchased by grocery store shoppers. In terms of fruit, most grocery stores offer bananas, apples, oranges, blackberries, raspberries, grapes, pineapples, cantaloupes, watermelons, and more; other grocery stores with larger produce selections might offer the listed fruits in addition to less common fruits, including mangoes, honeydews, starfruits, coconuts, and more. Depending on the grocery store, customers can purchase fruits in a few different ways. Some stores will charge a set amount per pound of fruit, and will weigh customers' fruit purchases and bill them accordingly; other stores will charge customers for each piece of fruit they buy, or for bundles of fruit (a bag of bananas, a bag of apples, etc.); other stores yet will simply charge by the container. Vegetables, including lettuce, corn, tomatoes, onions, celery, cucumbers, mushrooms, and more are also sold at many grocery stores, and are purchased similarly to the way that fruits are. Grocery stores typically stock more vegetables than fruit at any given time, as vegetables remain fresh longer than fruits do, generally speaking.}
    }%
}
    \caption{The dictated text that users recorded for the experiment.}
    \label{fig:text}
\end{figure}

Since Android and Apple smartphones have integrated a VA and represent approximately 99\% of the mobile market\footnote{\url{https://gs.statcounter.com/os-market-share/mobile/worldwide}} we opted to test only two VAs, Google Assistant and Siri. While others, e.g. Alexa and Bixby, might have millions of installations and dedicated devices, they might not be available to all potential participants and introduce representation biases. All participants had to enable the VA and then train it using their own voice to be eligible for the experiment and exposed to attacks. As described, the experiment's execution consisted of three phases. In the first phase, data collection, the participants were requested to provide one mobile recording and a video recorded from another device, e.g. a desktop computer. The mobile recording represents the face-to-face and spyware attacks, and the video recording represents the attack from published social media content. We requested the participants to record the samples in places with no ambient noise. All participants read the same text, which did not include commands that could cause the activation of the VA to increase the results' credibility. Additionally, the participants were requested to provide information about their mobile hardware and software versions to accommodate the quality of the statistical results. Using the files the participants sent, we synthesised the "attack" recordings with the phrase "Hey Google, call John" and "Hey Siri, call John" (depending on their devices) using an off-the-shelf open-source solution, namely the "Real Time Voice Cloning" project. In the second phase of the execution, we sent participants the synthesised voice samples and asked them to use them to attack their VAs, e.g. reproduce the content in proximity to the VA. Please note that we consider the smartphones unlocked. We argue that this is a soft requirement, as VAs in a smart home environment and not integrated into a smartphone device, so they would not have any type of lock. Finally, participants forwarded the results of the attacks and their observations for the last phase, which is the processing and presenting of the outcomes reported by the users.

\subsection{Dataset composition and overall findings}
\begin{figure}[th!]
    \centering
    \begin{tikzpicture}
    \pgfplotsset{width=.6\columnwidth,compat=1.8}
      \begin{axis}[
        ybar stacked, ymin=0,  
        bar width=25mm,
        enlarge x limits=0.4,
        legend columns=-1,
        legend style={draw=none},
        symbolic x coords={Male,Female},
        xtick=data,
        nodes near coords, 
        nodes near coords align={anchor=north},
        every node near coord/.style={},
      ]
      \addplot [myblue,fill=myblue,text=white] coordinates {
    ({Male},65)
    ({Female},23)};
      \addplot [myred,fill=myred,text=white] coordinates {
    ({Male},32)
    ({Female},20)};
      \legend{Android,iOS}
      \end{axis}
  \end{tikzpicture}

    \caption{Participant distribution per OS and gender.}
    \label{fig:part_distribution}
\end{figure}
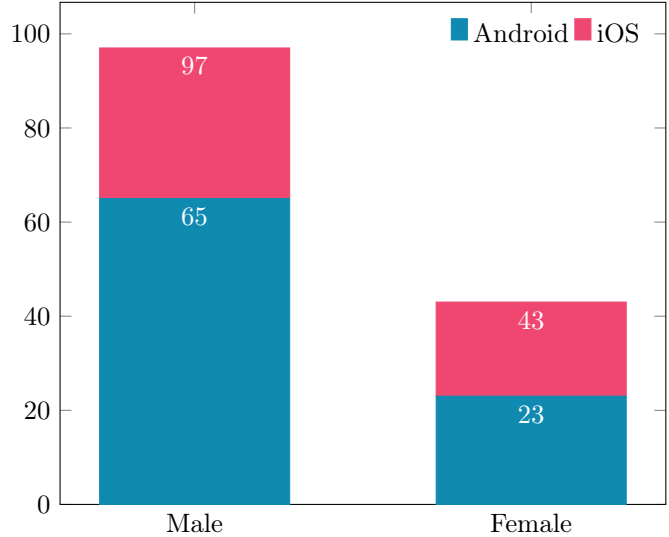

\begin{table}[th]
\begin{subtable}[h]{0.6\columnwidth}
    \centering
    \begin{tabular}{ll|ll}
    \hline
    \textbf{Manufacturer} & \# & \textbf{Version} & \textbf{\#} \\ \hline
    Xiaomi & 49 & 7 & 1 \\
    Samsung & 32 & 8 & 4 \\
    REALME & 3 & 9 & 4 \\
    Other & 4 & 10 & 9 \\
     &  & 11 & 37 \\
     &  & 12 & 30 \\
     &  & 13 & 3 \\ \hline
    \end{tabular}
    \caption{Android vendors and versions.}
    \label{tbl:android_vendors}
\end{subtable}
\begin{subtable}[h]{0.35\columnwidth}
    \centering
    \begin{tabular}{ll}
    \hline
    \textbf{Version} & \textbf{\#} \\ \hline
    14.8 & 1 \\
    15.5 & 1 \\
    15.6 & 5 \\
    15.7 & 4 \\
    16.0 & 19 \\
    16.1 & 16 \\
    16.2 & 5\\ 
    16.11 & 1\\ \hline
    \end{tabular}
    \caption{iOS versions.}
    \label{tbl:ios_versions}
\end{subtable}
\caption{Distribution of vendors and versions in our sample.}
\label{tbl:distr}
\end{table}

\
In total, 140 people participated in the experiment. The participants were not native English speakers and belonged to the age group of 18-40. In terms of the used operating system, 88 participants had Android devices, while the rest 52 had iOS. Figure \ref{fig:part_distribution} demonstrates the distribution of OS per gender, while Table \ref{tbl:android_vendors} shows the distribution of participants among vendors for Android and versions. For simplicity, we only report the major Android version. Similarly, for iOS, Table \ref{tbl:ios_versions} reports the iOS versions of the participants. For clarity, we report only the major release and brunch. Regarding vendors, the vast majority is shared by two companies, namely Xiaomi and Samsung. Therefore, our sample resembles the mobile vendor market share in Europe\footnote{\url{https://gs.statcounter.com/vendor-market-share/mobile/europe}}.

The participants performed multiple repetitions using the generated recordings. The results varied and could be separated into three categories: fully successful results, semi-successful results (noted as \textit{trigger}); when the VA was triggered but did not understand the command that was addressed, and finally unsuccessful where there was no response and action from the VA. From now on, we will refer to the attacks that the participants provided their voice samples, and we synthesised the voice commands as audio attacks. For the voice samples that the audio was extracted from the videos, we will refer to them as voice attacks. 

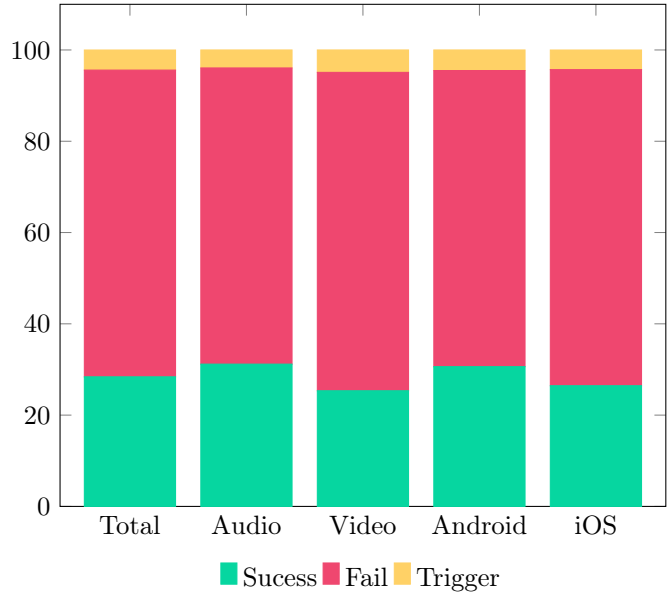
\begin{figure}[th!]
\centering
    \begin{tikzpicture}
    \pgfplotsset{width=.6\columnwidth,compat=1.8}
      \begin{axis}[
        ybar stacked, ymin=0,  
        bar width=12mm,
        enlarge x limits=0.15,
        legend columns=-1,
        legend style={draw=none,at={(0.5,-0.1)},anchor=north},
        symbolic x coords={Total,Audio,Video,Android,iOS},
        xtick=data,
        nodes near coords align={anchor=north},
        every node near coord/.style={},
      ]
      \addplot [mygreen,fill=mygreen] coordinates {
    ({Total},28.39)
    ({Audio},31.17)
    ({Video},25.34)
    ({Android},30.62)
    ({iOS},26.44)
    };
      \addplot [myred,fill=myred] coordinates {
    ({Total},67.2)
    ({Audio},64.89)
    ({Video},69.75)
    ({Android},64.87)
    ({iOS},69.25)
    };
    \addplot [myyellow,fill=myyellow] coordinates {
    ({Total},4.4)
    ({Audio},3.94)
    ({Video},4.91)
    ({Android},4.51)
    ({iOS},4.3)
    };
      \legend{Sucess,Fail,Trigger}
      \end{axis}
  \end{tikzpicture}
  \caption{Overview of experimental results.}
  \label{fig:all_results}
\end{figure}

In general, as illustrated in Figure \ref{fig:all_results}, we have come up with some consistent patterns. The overall success rate is on a scale of 3 out of 10 (28.39\%), with a significantly higher success rate (31.17\%) in audio attacks and a significantly lower success rate (25.34\%) in video attacks. The overall variation in trigger results is on the scale of 0.5\%, so it is considered insignificant. However, the success rate between the two OS platforms  differs statistically significantly (4.2\%).

More precisely, out of the 2180 attacks that were carried out, 619 were successful (28.39\%), 96 were triggers (4.4\%), and 1465 were unsuccessful (67.2\%). Figure \ref{fig:all_results} illustrates all these results. As highlighted, there is a higher success rate in audio attacks. More precisely, out of 1142 audio recordings, 356 were successful (31.17\%), and 741 failed (64.89\%). On the contrary, out of the 1038 video recordings, only 263 were successful (25.34\%), and 724 failed (69.75\%).

\section{Discussion}
In all cases, for the most part, positive results emerged, which raised many questions. If a VA can be tricked so easily, that is, with applications that are either available for free or easy to implement, how easy could it be to trick a security system where all the devices are connected to the same network? For example, a Smart Home consists entirely of such devices. What if anyone could open a person's home anytime with just one voice command? Are we heading into a modern age faster than we should, leaving huge security gaps behind? Over the years, more and more devices have become part of Smart Homes, trying to offer more convenience to users, yet this may come at a considerable cost.

\begin{figure}[th!]
\centering
    \begin{tikzpicture}
    \pgfplotsset{width=.6\columnwidth,compat=1.8}
      \begin{axis}[
        ybar stacked, ymin=0,  
        bar width=12mm,
        enlarge x limits=0.15,
        legend columns=-1,
        legend style={draw=none,at={(0.5,-0.1)},anchor=north},
        symbolic x coords={Android,Samsung,Xiaomi,iOS,Overall},
        xtick=data,
        nodes near coords align={anchor=north},
        every node near coord/.style={},
      ]
      \addplot [mygreen,fill=mygreen] coordinates {
    ({Android},34.40)
    ({Samsung},44.37)
    ({Xiaomi},32.96)
    ({iOS},28.36)
    ({Overall},31.17)
    };
      \addplot [myred,fill=myred] coordinates {
    ({Android},61.47)
    ({Samsung},54.23)
    ({Xiaomi},62.22)
    ({iOS},67.87)
    ({Overall},64.89)
    };
    \addplot [myyellow,fill=myyellow] coordinates {
    ({Android},4.14)
    ({Samsung},1.41)
    ({Xiaomi},4.81)
    ({iOS},3.77)
    ({Overall},3.94)
    };
      \legend{Sucess,Fail,Trigger}
      \end{axis}
  \end{tikzpicture}
  \caption{Audio attacks per vendor.}
  \label{fig:audio_attack_vendor}
\end{figure}
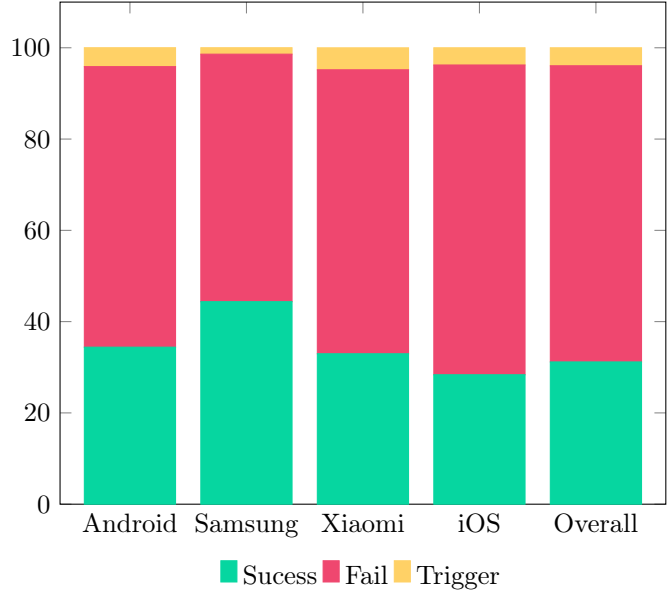

\begin{figure}[th!]
\centering
    \begin{tikzpicture}
    \pgfplotsset{width=.6\columnwidth,compat=1.8}
      \begin{axis}[
        ybar stacked, ymin=0,  
        bar width=12mm,
        enlarge x limits=0.15,
        legend columns=-1,
        legend style={draw=none,at={(0.5,-0.1)},anchor=north},
        symbolic x coords={Android,Samsung,Xiaomi,iOS,Overall},
        xtick=data,
        nodes near coords align={anchor=north},
        every node near coord/.style={},
      ]
      \addplot [mygreen,fill=mygreen] coordinates {
    ({Android},26.49)
    ({Samsung},25)
    ({Xiaomi},28.16)
    ({iOS},24.32)
    ({Overall},25.34)
    };
      \addplot [myred,fill=myred] coordinates {
    ({Android},68.58)
    ({Samsung},71.77)
    ({Xiaomi},67.35)
    ({iOS},70.78)
    ({Overall},69.75)
    };
    \addplot [myyellow,fill=myyellow] coordinates {
    ({Android},4.93)
    ({Samsung},3.23)
    ({Xiaomi},4.49)
    ({iOS},4.9)
    ({Overall},4.91)
    };
      \legend{Sucess,Fail,Trigger}
      \end{axis}
  \end{tikzpicture}
  \caption{Video attacks per vendor.}
  \label{fig:vid_attack_vendor}
\end{figure}
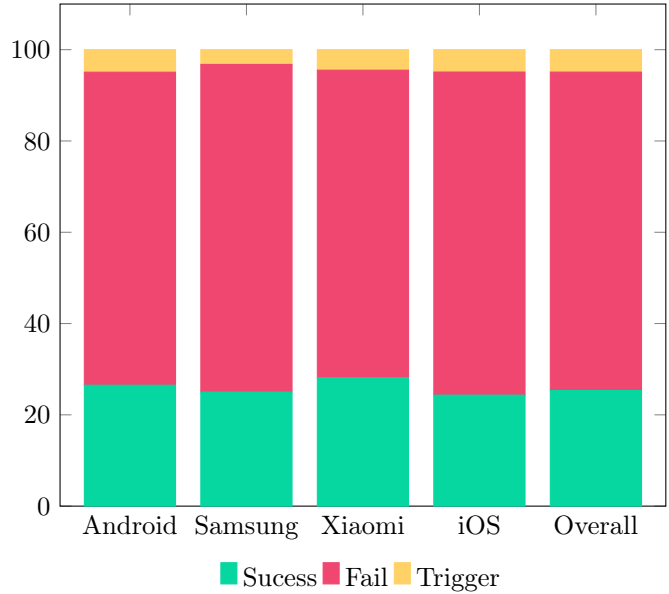

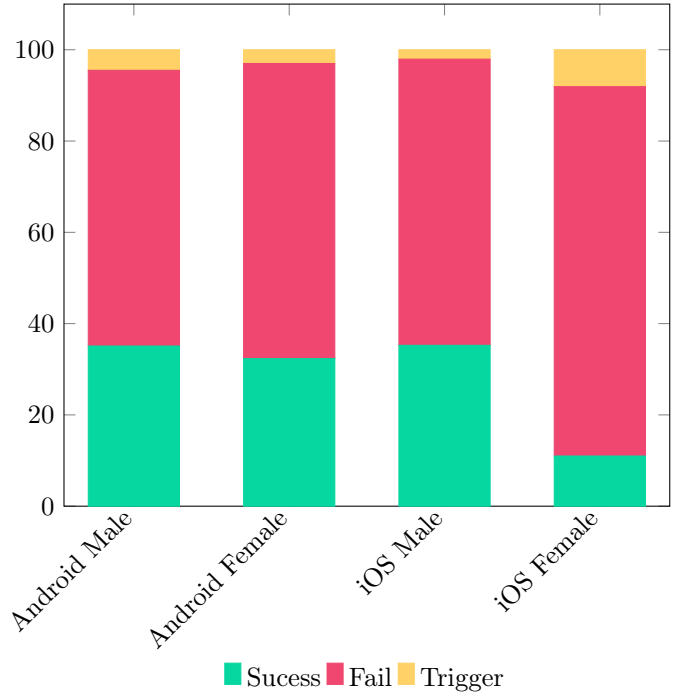
\begin{figure}[th!]
\centering
    \begin{tikzpicture}
    \pgfplotsset{width=.6\columnwidth,compat=1.8}
      \begin{axis}[
        ybar stacked, ymin=0,  
        bar width=12mm,
        enlarge x limits=0.15,
        legend columns=-1,
        legend style={draw=none,at={(0.5,-0.3)},anchor=north},
        symbolic x coords={AndroidM,AndroidF,iOSM,iOSF},
        xticklabels={Android Male, Android Female, iOS Male, iOS Female},
        xtick=data,
        x tick label style={rotate=45,anchor=east},
        nodes near coords align={anchor=north},
        every node near coord/.style={},
      ]
      \addplot [mygreen,fill=mygreen] coordinates {
    ({AndroidM},35.09)
({AndroidF},32.33)
    ({iOSM},35.24)
    ({iOSF},10.99)
    };
      \addplot [myred,fill=myred] coordinates {
 ({AndroidM},60.4)
({AndroidF},64.66)
    ({iOSM},62.7)
    ({iOSF},80.92)
    };
    \addplot [myyellow,fill=myyellow] coordinates {
({AndroidM},4.51)
({AndroidF},3.01)
    ({iOSM},2.06)
    ({iOSF},8.09)
    };
      \legend{Sucess,Fail,Trigger}
      \end{axis}
  \end{tikzpicture}
  \caption{Audio attacks per OS and Gender.}
  \label{fig:aud_attack_gender}
\end{figure}

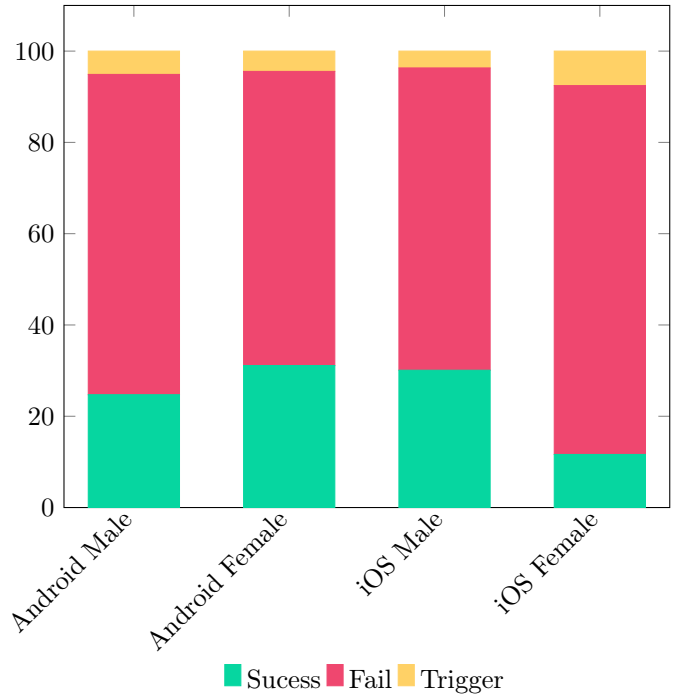
\begin{figure}[th!]
\centering
    \begin{tikzpicture}
    \pgfplotsset{width=.6\columnwidth,compat=1.8}
      \begin{axis}[
        ybar stacked, ymin=0,  
        bar width=12mm,
        enlarge x limits=0.15,
        legend columns=-1,
        legend style={draw=none,at={(0.5,-0.3)},anchor=north},
        symbolic x coords={AndroidM,AndroidF,iOSM,iOSF},
        xtick=data,
        xticklabels={Android Male, Android Female, iOS Male, iOS Female},
        x tick label style={rotate=45,anchor=east},
        nodes near coords align={anchor=north},
        every node near coord/.style={},
      ]
      \addplot [mygreen,fill=mygreen] coordinates {
    ({AndroidM},24.72)
({AndroidF},31.11)
    ({iOSM},30.08)
    ({iOSF},11.63)
    };
      \addplot [myred,fill=myred] coordinates {
 ({AndroidM},70.17)
({AndroidF},64.44)
    ({iOSM},66.23)
    ({iOSF},80.81)
    };
    \addplot [myyellow,fill=myyellow] coordinates {
({AndroidM},5.11)
({AndroidF},4.45)
    ({iOSM},3.69)
    ({iOSF},7.56)
    };
      \legend{Sucess,Fail,Trigger}
      \end{axis}
  \end{tikzpicture}
  \caption{Video attacks per OS and Gender.}
  \label{fig:vid_attack_gender}
\end{figure}

Human traits have been known to play a role in cybersecurity \cite{gratian2018correlating}; nevertheless, this role may not always be obvious. This can be augmented by biases in artificial intelligence \cite{osoba2017intelligence,ntoutsi2020bias} since, over the past few years, the convergence of cybersecurity and artificial intelligence is continuously growing to address the challenges posed by big data. Gender biases are often in artificial intelligence \cite{leavy2018gender} and may imply further issues for cybersecurity. Our work illustrates that there is a significant gender bias for attacks against iOS devices. Given that the samples of the participants in our experiment were processed exactly the same for both OSes, it is evident that iOS devices perceive the attacks entirely differently based on the gender of the user. This difference is so evident that the attacks against females were 10.98\% successful and 35.24\% against males, which is more than three times increase. Notably, the gender imbalance was also exhibited in the experiments of \cite{wenger2021hello}, who categorise gender as a decisive factor for the success of their attacks. Nevertheless, in our experiments, attacks against females were always less successful. The above illustrates that there inherent biases in the training of voice authentication systems which may trigger differently.

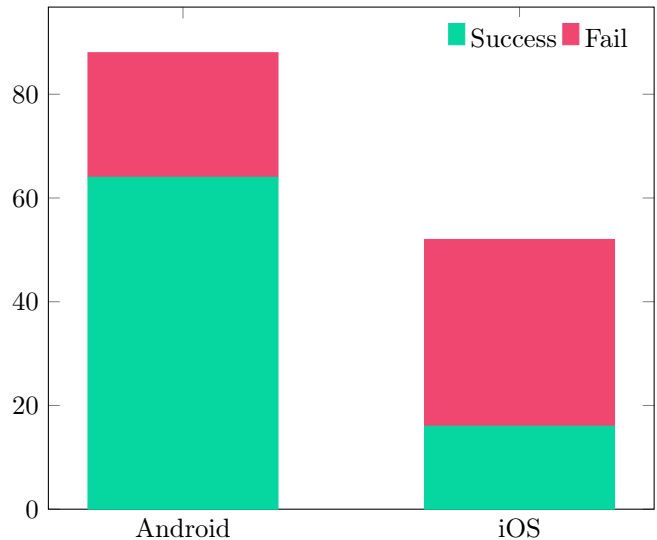
\begin{figure}
    \centering
 \begin{tikzpicture}
    \pgfplotsset{width=.6\columnwidth,compat=1.8}
      \begin{axis}[
        ybar stacked, ymin=0,  
        bar width=25mm,
        enlarge x limits=0.4,
        legend columns=-1,
        legend style={draw=none},
        symbolic x coords={Android,iOS},
        xtick=data,
        nodes near coords, 
        nodes near coords align={anchor=north},
        every node near coord/.style={},
      ]
      \addplot [mygreen,fill=mygreen] coordinates {
    ({Android},64)
    ({iOS},16)};
      \addplot [myred,fill=myred] coordinates {
    ({Android},24)
    ({iOS},36)};
      \legend{Success,Fail}
      \end{axis}
  \end{tikzpicture}
\caption{At least one successful attack per OS.}
    \label{fig:one_success}
\end{figure}

Finally, in Figure \ref{fig:one_success}, we illustrate the number of participants for which there was at least one successful attack. Given her proximity to the VA and lack of monitoring, an adversary may persistently perform attacks until she has the desired result. Again we notice that there are significant differences among the vendors. More precisely, for Android, for 72.73\% of the users, there is at least one successful attack, while for iOS, this drops down to 28.84\%. Practically, Android users are approximately 2.5 times more vulnerable than iOS users.

Having our presented results in mind, we should consider how much they could be improved if we incorporated even more sophisticated AI-empowered voice synthesis modules, which could provide voice outputs even closer to the real user ones.

\section{Conclusions}
The continuous integration of DAs enables seamless human-computer interaction. Even more, the use of VAs provides this functionality in a more human way. It is easier and more direct for us to ask for something we want by speaking. This convenience comes with several drawbacks, as VAs are not bulletproof. On the contrary, many attacks in the literature illustrate various ways to trick VAs into performing tasks that are not initiated by the users. Following this line of research, we performed a large-scale experiment, the largest in the related work, to assess the security of two of the most widely used commercial VAs against synthesised voices. One of the key differences is that the VAs were trained with the users' voices so that they would not be triggered to perform any task with their "master's" voice.

Our experiments illustrate some startling facts. More precisely, in all VAs using an open-source voice synthesiser with voice samples provided by the participants, approximately one out of three attacks is successful. Even more, there is at least one successful attack for more than half of the participants. Our experiments indicate that there are big variations among vendors regarding their susceptibility to such attacks. Additionally, there are underlying gender biases that make the attacks significantly more robust for males in the case of iOS. 

We believe that the above requires more attention from manufacturers. The diversity of the results in terms of gender and manufacturer is subject to different interpretations. Yet, we believe that for systems that are so widely used and integrated into millions of devices and interconnected to so many others, such issues are very grave. For starters, given that most of these attacks would initially be prerecorded, a randomised request for response could be considered a temporary patch. In the long run, VAs should integrate audio deepfake detection mechanisms \cite{blue2022you,9975381} to allow VAs to determine whether a voice has been synthesised. Other additional measures may include the occasional use of 2FA or the proximity to a user device with the trusted voice for cases where sensitive content is requested, monetary transaction, or potentially dangerous command is detected.

\section*{Acknowledgments}
This work was supported by the European Commission under the Horizon Europe Programme, as part of the project LAZARUS (\url{https://lazarus-he.eu/}) (Grant Agreement no. 101070303).

The content of this article does not reflect the official opinion of the European Union. Responsibility for the information and views expressed therein lies entirely with the authors.

\bibliographystyle{IEEEtran}
\bibliography{refs}
\end{document}